\documentstyle[prl,twocolumn,aps]{revtex}
\input{epsf}
\begin{document}
\draft
\twocolumn[\hsize\textwidth\columnwidth\hsize\csname @twocolumnfalse\endcsname
\title{Quantum Computing of Classical Chaos: Smile of the Arnold-Schr\"odinger 
Cat}

\author{B. Georgeot and D. L. Shepelyansky}

\address {Laboratoire de Physique Quantique, UMR 5626 du CNRS, 
Universit\'e Paul Sabatier, F-31062 Toulouse Cedex 4, France}

\date{January 1, 2001}

\maketitle

\begin{abstract}
 We show on the example of the 
Arnold cat map that classical chaotic systems
can be simulated with exponential efficiency on a quantum computer.  Although
classical computer errors grow exponentially with time, the quantum algorithm
with moderate imperfections is able to simulate accurately the unstable chaotic
classical dynamics for long times.  The algorithm can be easily implemented
on systems of a few qubits.
\end{abstract}
\vskip1pc]

\narrowtext

A great deal of attention has been attracted recently by the possibility
to perform numerical simulations on a quantum computer. The massive
parallelism allowed by quantum mechanics enables to operate on an exponential
number of states using a single quantum transformation, as was stressed by
Feynman \cite{feynman}.  However, even if exponential gain may be
possible in such quantum simulations, compared to the computations
on classical computers, only few problems have been found where
an explicit quantum algorithm displays such efficiency.  The most famous 
of them is the factorization of large integers, for which Shor \cite{shor1}
constructed an explicit algorithm which is exponentially faster than any
known classical algorithm.  Another well-known algorithm, invented 
by Grover \cite{grover}, also shows that quantum mechanics can enormously 
accelerate the search problem in an unsorted database, although the gain is
not exponential.  Although quantum-mechanical problems are computationally
very hard for classical simulations, at present only few physical
systems are known which can be simulated with exponential efficiency on a
quantum computer.  Such systems include certain spin lattices \cite{molmer},
some types of many-body systems \cite{lloyd}, and since recently the kicked
rotator model of quantum chaos \cite{qckr}.  The advances in the field of
quantum computation \cite{divincenzo,josza,steane} generated many proposals for the
experimental realization of such a computer. This computer is viewed as
a system of qubits (two-level systems) on which one-qubit rotations and two-qubit
transformations allow to realize any unitary transformation in the exponentially
large Hilbert space (see reviews \cite{divincenzo,josza,steane}).
 At present operations with 
two qubits were realized with cold ions \cite{monroe}, and the Grover algorithm 
was performed on a three-qubit system built on nuclear spins in a molecule 
\cite{trois}.

It may seem natural that quantum computers can simulate efficiently the
evolution of certain quantum systems.  Such systems are very hard to
simulate on classical computers due to the exponentially large Hilbert space.
However, there also exists a large class of classical Hamiltonian systems
which are very hard to simulate accurately on a classical computer.  Indeed,
the systems displaying dynamical chaos are characterized by an exponential
local instability of trajectories in the phase space \cite{arnold,lieberman}. 
 As a result, standard
round-off errors of an usual computer grow exponentially with time, and give
a complete change of a dynamical trajectory with given initial parameters after
a few characteristic periods of the system motion.  In this situation, 
the simulation of a full phase space density even for moderate times needs
an exponential number of orbits and soon exceeds the capacity of modern classical
computers. To our knowledge, the problem of performing such simulations on a
quantum computer was not addressed until now.  Indeed, it may look surprising
that quantum mechanics may help in simulations of classical dynamics.  In this
paper, we show that a well-known example of classical chaotic system can
be simulated on a quantum computer with exponential efficiency compared to
classical algorithms.  Moreover, even if due to chaos 
the classical errors grow exponentially with time, the quantum simulations with
moderate quantum errors still enable to reproduce accurately the time evolution
in the classical phase space. The resolution of this apparent paradox is
rooted in the fundamental differences between classical and quantum mechanics.

One of the most famous example of classically chaotic systems is the Arnold 
cat map, an automorphism of the torus \cite{arnold,lieberman}.  The dynamics of the map is given by:
\begin{equation}
\label{catmap}
\bar{y}=y+x \; \mbox{(mod} \;\mbox{1)}\;\;, \;\; \bar{x}=y+2x \;\mbox{(mod} 
\;\mbox{1)}\;,
\end{equation}
where bars denote the new values of the variables after one iteration.  This 
is an area-preserving map, in which $x$ can be considered as the space variable
and $y$ as the momentum. In this way, the first equation can be seen
as a kick which changes the momentum $y$, while the second equation describes
the free phase rotation.  This map belongs to the class of Anosov systems, 
with homogeneous exponential divergence of trajectories and positive 
Kolmogorov-Sinai entropy $h=\mbox{ln} (\frac{3+\sqrt{5}}{2}) \approx 0.96$.
Due to this exponential instability, a typical computer round-off error of
order $10^{-16}$ will change completely the position of a trajectory in the
phase space torus after only 38 iterations.  Although the exact dynamics of 
(\ref{catmap}) is time-reversible \cite{note1}, the round-off errors make
it effectively irreversible after a short time.

Usually computer round-off errors are not symplectic, and destroy the 
area-preserving property of the map.  However, it is possible to consider
a discretized map which remains area-preserving after discretization. It
is known that such symplectic discretization describes the continuous
dynamics in the most appropriate way \cite{symplectic}.  Such a discrete
approximation remains close to the exact map dynamics up to the time scale
$t_E \approx \ln N/h$ where $N^2$ is the number of points of the discretized 
torus, so that the discrete cells have the area $1/N^2$.  For the cat map 
(\ref{catmap}) the discretized map is especially simple, consisting of
the dynamics through (\ref{catmap}) of the $N^2$ points $(x_i,y_j)$ with 
$x_i=i/N$, $i=0,...,N-1$ and $y_j=j/N$, $j=0,...,N-1$. Even after discretization, 
the exponential instability still manifests itself through the rapid 
disappearance of any structure in phase space (for example the cat image)
after few iterations, see
Fig.1 (left). The discrete map
preserves time-reversibility \cite{note1}, however any small imprecision at the
time of inversion destroys this reversibility as is illustrated on Fig.1 (left).
Here for $N=128$, the smallest error (of one cell size) destroys reversibility
already after 10 iterations. On a classical computer, one map iteration 
requires $O(N^2)$ additions to simulate the evolution of a phase-space density
distribution.

On the contrary, we found that on a quantum computer the discretized Arnold cat
map can be simulated exponentially faster.
  Our quantum algorithm operates on $3n_q -1$ qubits. The first two
quantum registers, each with $n_q$
qubits, describe the position $x_i$ and the momentum $y_j$ of $N^2$ points 
of the discretized classical phase space,
with $N=2^{n_q}$. The remaining $n_q-1$ qubits are used as workspace.  
An initial classical phase space density can then be represented by a
quantum state $\sum_{i,j} a_{ij} |x_i> |y_j>|0>$.
 The map
dynamics requires additions of integers modulo (N) (modular additions).  
The quantum algorithm we use for this operation is similar to the one described
in \cite{barrenco} (see also \cite{preskill}). The third register holds the carries 
of the addition, and the result is taken modulo (N) by eliminating
the last carry.  One map iteration  requires first adding the first register
to the second, and then adding the second register to the first. After
that, the coefficients $a_{ij}$ describe the classical phase space density
after one map iteration. 
To perform these additions, $8n_q -12$ Toffoli gates and $8n_q-10$
 controlled-not gates are needed per map iteration, 
giving a total of $16n_q-22$ operations for $n_q \geq 3$ \cite{note2}.
This means that the quantum computer can iterate this classical chaotic map 
exponentially faster than the classical computer, which requires $O(2^{2n_q})$
operations per iteration.  Hence, the quantum evolution obeying the 
Schr\"odinger equation describes the classical Arnold cat map, and we will call
this quantum dynamics the Arnold-Schr\"odinger cat map.

If the quantum gates are perfect, then the quantum algorithm describes
exactly the classical density evolution.  But physical systems are never
perfect, and to be really efficient the quantum algorithm should be stable
against 

\begin{figure}
\epsfxsize=3.4in
\epsfysize=6.8in
\epsffile{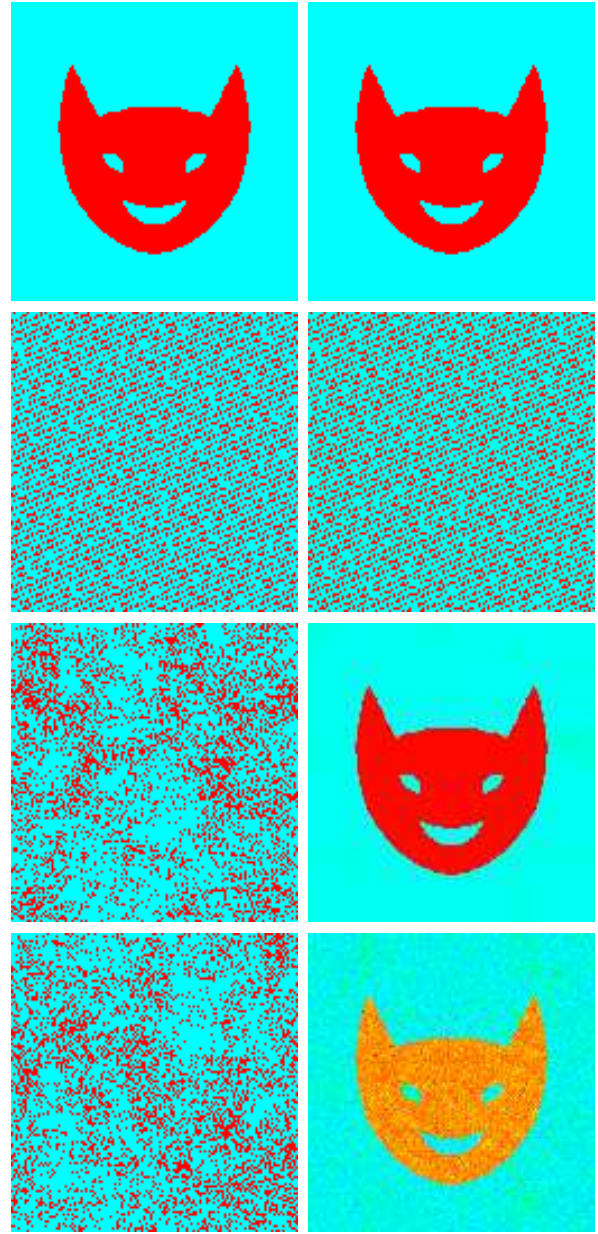}
\vglue 0.2cm
\caption{Dynamics of Arnold-Schr\"odinger cat simulated on a classical
(left) and quantum computer (right), on a $128 \times 128$ lattice.  Upper row: initial
distribution; second row: distributions after 10 iterations; third row: 
distributions at $t_{2r}=20$, with time inversion made at $t_r=10$; bottom row:
distributions at $t_{2r}=400$, with time inversion made at $t_r=200$. Left:
inversion is done with classical error of one cell size ($\epsilon=1/128$)
at $t=t_r$ only;
right: all quantum gates operate with quantum errors of amplitude 
$\epsilon =0.01$; 
color from blue to red gives the probability $|a_{ij}|^2$; $n_q=7$.
} 
\label{fig1}
\end{figure}

\noindent  imperfections. In view of the exponential instability of classical
computer errors in this problem, this may look rather doubtful.  To study
the effects of imperfections on this algorithm, we introduced some random 
unitary
noise in the gate operations.  For each gate transformation, the nondiagonal 
part was diagonalized, and each eigenvalue was multiplied by a random phase 
$\exp(i\eta)$, with $-\epsilon <\eta < \epsilon$.  Here we assume that 
imperfections due to residual static coupling between qubits are small enough,
and that the quantum computer operates below the quantum chaos border
discussed in \cite{melting}.

\begin{figure}
\epsfxsize=3.4in
\epsfysize=2.0in
\epsffile{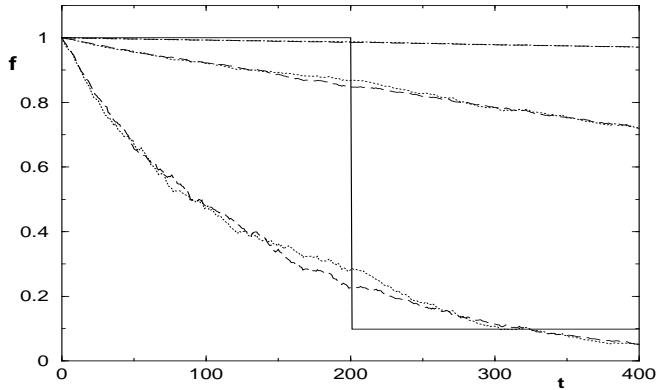}
\vglue 0.2cm
\caption{Quantum fidelity $f$ of Arnold-Schr\"odinger cat as a function of
 time $t$ for quantum errors $\epsilon = 0.003, 0.01, 0.03$
(dashed and dotted curves from top to bottom respectively). 
Initial state: cat's smile as in Fig.~1 (dashed curves)
and line $x=1/2$ (dotted curves). Full curve shows 
the drop of fidelity when a minimal classical error is done at $t=200$ 
(see text).} 
\label{fig2}
\end{figure}
\vglue -0.6cm
\begin{figure}
\epsfxsize=3.4in
\epsfysize=1.8in
\epsffile{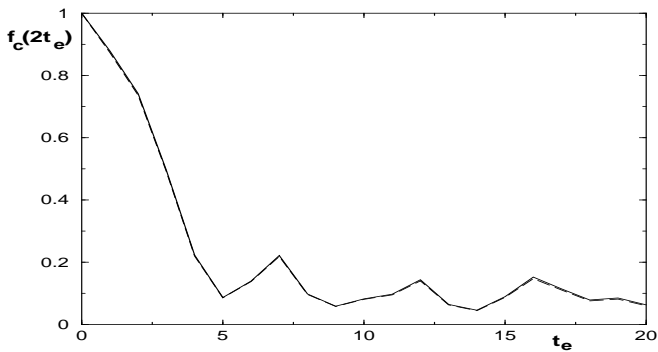}
\vglue 0.2cm
\caption{Classical fidelity $f_c(2t_e)$ vs. time $t_e$ 
when the minimal classical error ($\epsilon =1/128$) is made (full curve).
Dashed curve shows the same $f_c$ obtained by the quantum computer
with imperfections of amplitude $\epsilon=0.01$ (see text).} 
\label{fig3}
\end{figure}

To investigate the stability of the algorithm with respect to quantum 
imperfections, we used first a time-inversion test.  Namely, starting from a 
given classical density (representing the cat's smile), we perform $t_r$ 
iterations forwards, 
then invert all momenta (time-inversion), and perform again $t_r$ iterations.
Without imperfections, the density returns exactly to its initial distribution
at $t_{2r}=2t_r$.
On the left of Fig.1, one can see the dramatic effect of small random
classical computer errors (here of size $\epsilon=1/N \approx 10^{-2}$), 
performed only at the moment of the time-inversion: it
completely destroys reversibility after a few iterations.  On the contrary, 
the quantum errors of similar amplitude, although present at each map 
iteration, practically do not affect the smile of the Arnold-Schr\"odinger cat
after $t_{2r}=20$ iterations, and only slightly perturb it after $t_{2r}=400$.
This pictorial image shows the power of quantum computation, which even in
 presence of relatively strong imperfections is able to simulate classical
chaotic dynamics.  We note that quantum systems for which the classical limit 
is chaotic (e.g. the kicked rotator) are also stable with respect to time 
inversion \cite{dima83}.

To be more quantitative, we computed the fidelity of the quantum state
in presence of errors, namely $f(t)=|<\psi_{\epsilon}(t)|\psi_0 (t)>|^2$.
Here $|\psi_0 (t)>$ is the quantum state after $t$ perfect iterations, while
$|\psi_{\epsilon}(t)>$ is the quantum state after $t$ imperfect iterations.
The dependence of fidelity on time is shown on Fig.2. 
Here we present $f(t)$ for two initial states, one representing the
cat's smile, and another a line in phase space, with $x=1/2$.  The latter
is especially easy to prepare, requiring only $n_q+1$ single-qubit rotations.
 The data clearly
show that in both cases the fidelity drops very slowly with the number of 
iterations, confirming the stability of quantum dynamics.  In view of the
exponential growth of classical errors, this may look as a paradox. Indeed,
as is illustrated in Fig.1, exponentially small classical errors of size
$1/N$ destroy practically immediately any structure. The
resolution of this paradox lies in the fact that a small classical error 
can be very large from the viewpoint of quantum mechanics. This fact is shown
in Fig.2, where after a small classical error affecting only the smallest bit
in the positions $x_i,y_j$ the fidelity of the quantum state drops immediately
to a very small value.  Curiously enough, after this drop, perfect iterations
of the map do not change the fidelity, although the classical error (i.e.
distance between exact and perturbed orbits)  starts
to grow exponentially due to trajectory divergence in phase space.

In this situation, one may wonder where in the quantum dynamics 
 is hidden the classical exponential instability.  In fact, it
is always present even if quantum dynamics remains stable.  Indeed, the drop
in the fidelity induced by classical errors depends exponentially on the
moment of time $t_e$ when the error is made.  This fact is illustrated by
Fig.3, which shows the classical fidelity $f_c$, defined
in the same way as the fidelity for quantum errors:  
$f_c (t,t_e)=|<\psi_{e}(t,t_e)|\psi_0(t)>|^2$ where $|\psi_{e}(t,t_e)>$ is the 
quantum state after the classical error is done at time $t_e < t$ and
 $|\psi_0 (t)>$ is the quantum state without error.  
This function $f_c (t,t_e)$ can be also computed purely classically.
As seen in Fig.2, for given $t_e$ and $t>t_e$, the fidelity $f_c$ remains
exactly constant, since $f_c$ is preserved by unitary transformations.  
However, its value depends strongly on $t_e$, as is shown in Fig.3, where
$f_c$ is computed at time $t=2t_e$.  The data
clearly show the exponential drop of classical fidelity with $t_e$.  This
reflects the existence of exponential instability in the cat map dynamics.
If a time inversion with errors is done at $t_r=t_e$, as in Fig.3,
then the value of $f_c$
gives the recovered fraction of the initial distribution (cat's smile) 
at $t=2t_r$.
This whole process can be made on the quantum computer with imperfections, 
and Fig.3 shows that even with imperfections the quantum computer gives
practically the same classical fidelity which drops exponentially.  Hence
a quantum computer can simulate accurately the exponential growth of classical 
errors in the regime of chaos.

\begin{figure}
\epsfxsize=3.4in
\epsfysize=2.0in
\epsffile{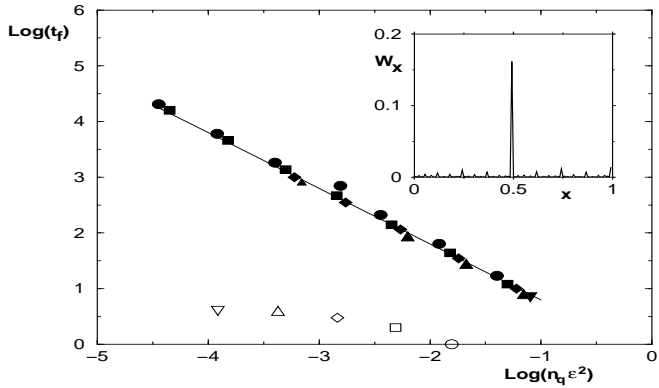}
\vglue 0.2cm
\caption{Fidelity time scale $t_f$ as a function of $\epsilon^2 n_q$:
$n_q=4$ (circles), $5$ (squares), $6$(diamonds), $7$ (triangles up), 
$8$ (triangles down)); filled symbols are for quantum
errors ($0.003 \leq \epsilon \leq 0.1$), open ones are for classical errors
($0.003 < \epsilon < 0.1$); the full line gives $t_f=0.63/(\epsilon^2n_q)$.
Inset: probability distribution $W_x$ in $|x>$ at the moment of return $t_{2r}=400$
for time inversion at $t_r=200$, and quantum imperfections $\epsilon=0.03$,
for $n_q=7$ with $x=1/2$ at $t=0$.} 
\label{fig4}
\end{figure}

To study quantitatively the dependence of the fidelity $f(t)$ on the magnitude
of errors, we determine the fidelity time scale $t_f$ by the condition
$f(t_f)=0.5$.  For quantum errors, Fig.4 shows that 
$t_f \approx 0.63/(\epsilon^2n_q)$.  Indeed, the probability of transition 
from the exact state to other states is of order $\epsilon^2$ for each gate operation.
After $t_f$ map iterations, $t_f n_q$ such operations are done, so that 
the fidelity drops by $\epsilon^2 n_q t_f \sim 0.5$, giving the above estimate.
This estimate is rather general, and it corresponds to a general
property of quantum mechanics due to which 
the fidelity can drop only polynomially with unitary noise and
the number of imperfect gates applied.  On the contrary, in the classical
case $t_f$ extracted from the classical fidelity $f_c(t_e=t_f)=0.5$ (see Fig.3)
is of the order of $t_f \approx 1.4 \ln (1/\epsilon)$, comparable with $t_E$ for
$\epsilon \sim 1/N$.

We stress that the
Arnold-Schr\"odinger cat is very simple to implement.  For example,
one map iteration with $n_q=4$ requires only $11$ qubits and $42$ gates,
and can be experimentally realized in the near future.  The time inversion
test explained above can be performed experimentally and be used to test 
the actual accuracy of the quantum computer.  Indeed, an initial distribution
in the form of the line $x=1/2$ can be easily prepared, and from a few 
measurements of the $|x>$ register at the return moment $t=t_{2r}$ 
one can estimate the probability of non-return which allows to determine
the amplitude of quantum errors. The inset in Fig.4 shows an example of
such final state.  It is interesting to note that $n_q=20$
needs only $59$ qubits and will permit to make computations unaccessible
to nowadays supercomputers, with memory size $\approx 200$ Go.  
In this regime global
quantities inaccessible by classical computation can be obtained.  For example,
the main harmonics of the density distribution can be obtained with the help
 of the quantum Fourier transform followed by a few measurements. 

In conclusion, our study of the Arnold-Schr\"odinger cat dynamics shows that
classical unstable motion, for which  classical
computers demonstrate exponential sensibility to errors, 
can be simulated accurately with exponential efficiency by a realistic
quantum computer. 

We thank the IDRIS in Orsay and the CalMiP in Toulouse for access to 
their supercomputers.


\begin{thebibliography}{99}
\bibitem{note1} The time-inversion physically can be viewed as inversion 
of all momenta $y \rightarrow -y$ half-way between the kicks.  The iteration
with time-inversion gives $\bar{y} = -y -x $, $\bar{x}=x$.  After that,
iterations of (\ref{catmap}) give the evolution backwards.
\bibitem{note2} The time-inversion in the quantum algorithm requires $n_q+1$
one-qubit rotation ($|0> \rightarrow |1>$, $|1> \rightarrow |0>$),  plus
$10n_q-11$ Toffoli and controlled-not gates, and hence 
is also exponentially fast.
\end{thebibliography}
\end{document}